\documentclass{edp-conf}
\usepackage{graphicx}
\usepackage{floatflt}

\begin{document}

\TitreGlobal{SF2A 2003}

\title{The X-ray flaring activity of the galactic nucleus observed with XMM-Newton}
\author{E.~Brion}\address{CENBG, Domaine du Haut-Vigneau, BP 120, 33175 Gradignan Cedex, France\\
E-mail: \texttt{brion@cenbg.in2p3.fr}}
\author{A.~Goldwurm}\address{CEA/DSM/DAPNIA/SAp, CE Saclay, 91191 Gif-sur-Yvette Cedex, France}
\author{P.~Goldoni$^2$}
\author{P.~Ferrando$^2$}
\author{F.~Daigne}\address{IAP, 98 bis boulevard Arago, 75014 Paris, France}
\author{A.~Decourchelle$^2$}
\runningtitle{The X-ray flaring activity of Sgr~A* observed with XMM-Newton}
\setcounter{page}{237}
\index{Brion, E.}
\index{Goldwurm, A.}
\index{Goldoni, P.}
\index{Ferrando, P.}
\index{Daigne, F.}
\index{Decourchelle, A.}

\maketitle

\begin{abstract}
We report the results of  XMM-Newton observations of Sgr~A*, the radiative counterpart of the massive black hole at the nucleus of our Galaxy, performed in the frame of the guaranteed time survey program of the Galactic Center region. The discovery of bright X-ray flares from Sgr~A* with Chandra in October 2000 have opened new perspectives to understand the processes at work in this object and in general in black holes accreting at low accretion rates. We report here the important results obtained with XMM-Newton on the Sgr~A* high-energy flaring activity and we discuss the implications on the models and the future observational perspectives.
\end{abstract}

%

\section{Introduction}

Infrared (IR) observations on star proper motions near the Galactic Center have provided a compelling evidence for the presence of a massive Black Hole (BH) at the galactic nucleus. Coincident with the dynamical center, the bright ($\sim1$~Jy), variable, compact, synchrotron radio source Sgr~A* appears to be the counterpart of this object. Its radio spectrum is described by an inverted or flat power law with high and low frequency cut-offs, and a peculiar sub-mm bump at frequencies $>$~100~GHz. Only recently detected in IR at $3.8~\mu$m (Ghez et al. 2003), this source is undetectable in visible and UV due to the large absorption and is also very weak in X-rays. The Chandra Observatory measured in 1999 a quiescent luminosity of only $L_\mathrm{X}\mathrm{[2-10\ keV]}\approx2\times10^{33}$~erg~s$^{-1}$ from Sgr~A* and then detected in October 2000 a bright 3~h flare, characterized by rapid variability (Baganoff et al. 2001, 2003). The luminosity increased to $L_\mathrm{X} \mathrm{[2-10\ keV]}\approx10^{35}$~erg~s$^{-1}$ in 4000~s, the power law spectrum hardened, with a change of the photon index from  2.7 (quiescent phase) to an index of 1.3, and a rapid decrease on a timescale of 600~s was observed, implying an emitting region size $<20$~Schwarzschild radii ($R_\mathrm{S}$). We report the detection of another X-ray flare from Sgr A* observed with XMM-Newton on 4$^{\rm th}$ September 2001 (Goldwurm et al. 2003).

\section{Observations and results}

XMM-Newton was pointed towards the galactic nucleus for about 26~ks. The EPIC MOS and PN cameras were used in standard \textit{Full Frame} imaging mode with the medium filter. The count rate average in the central CCD ($11'\times11'$) combining the two MOS cameras was about 4.50~cts~s$^{-1}$. The recorded image of this complex region is dominated by the diffuse emission of the Sgr~A East region. Point sources can also be distinguished and an excess around Sgr~A* clearly present.

\begin{floatingfigure}[r]{6cm}
   \centering \includegraphics[width=5.9cm]{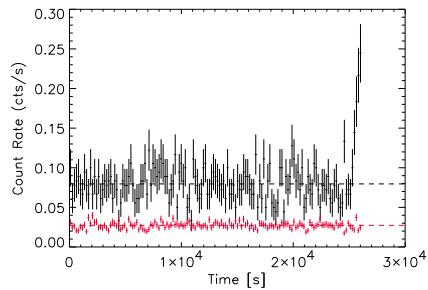}
   \caption{The $2-10$~keV count rate collected with both MOS cameras from the region within $10''$ from Sgr~A* (black upper curve) compared to the one from different region of the CCD (red lower curve).}
   \label{fig:lightcurve}
\end{floatingfigure}

\begin{figure}
\begin{center}
   \includegraphics[width=3.5cm]{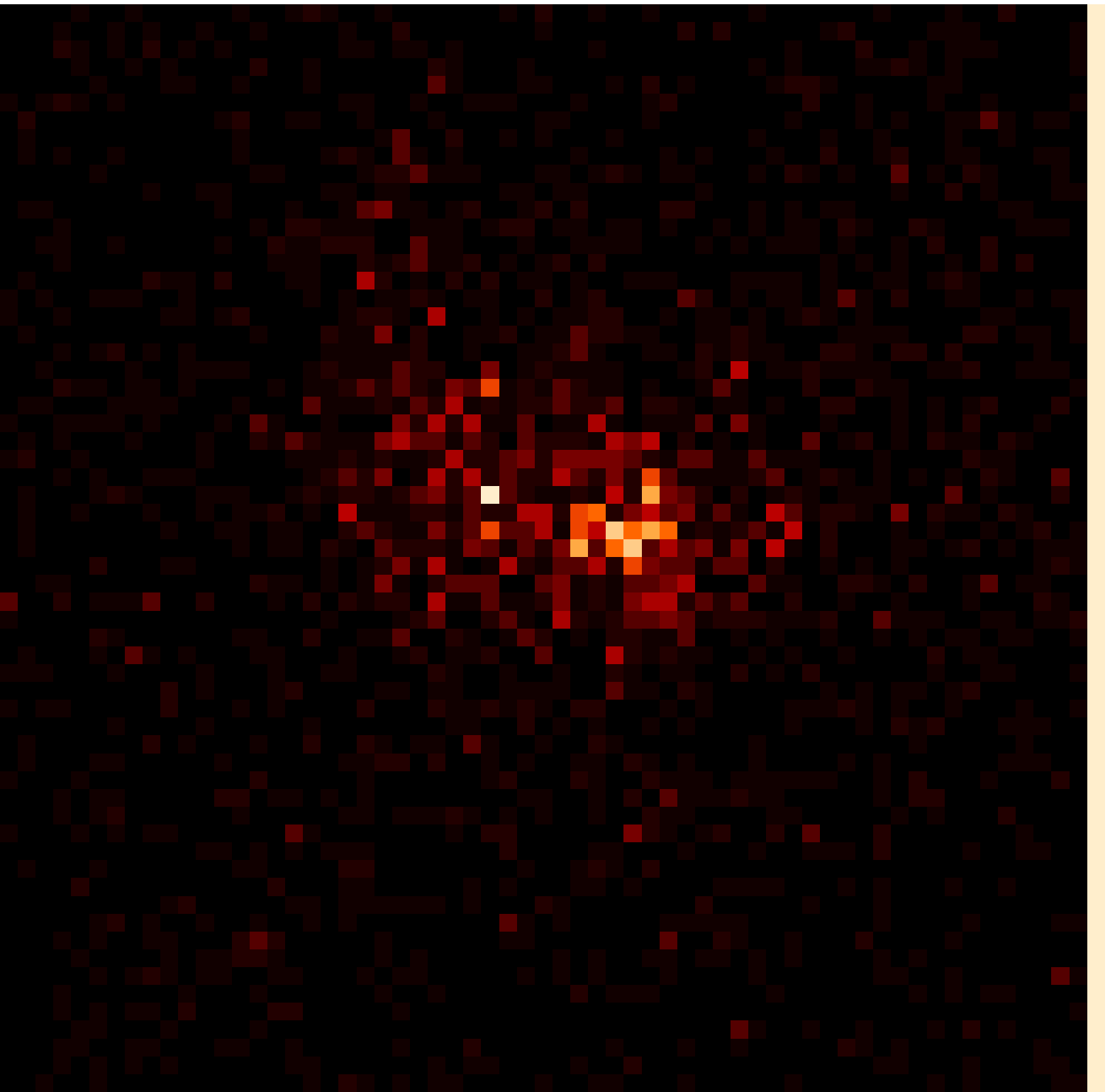}
   ~~~~~~~~~~~~~~~~
   \includegraphics[width=3.5cm]{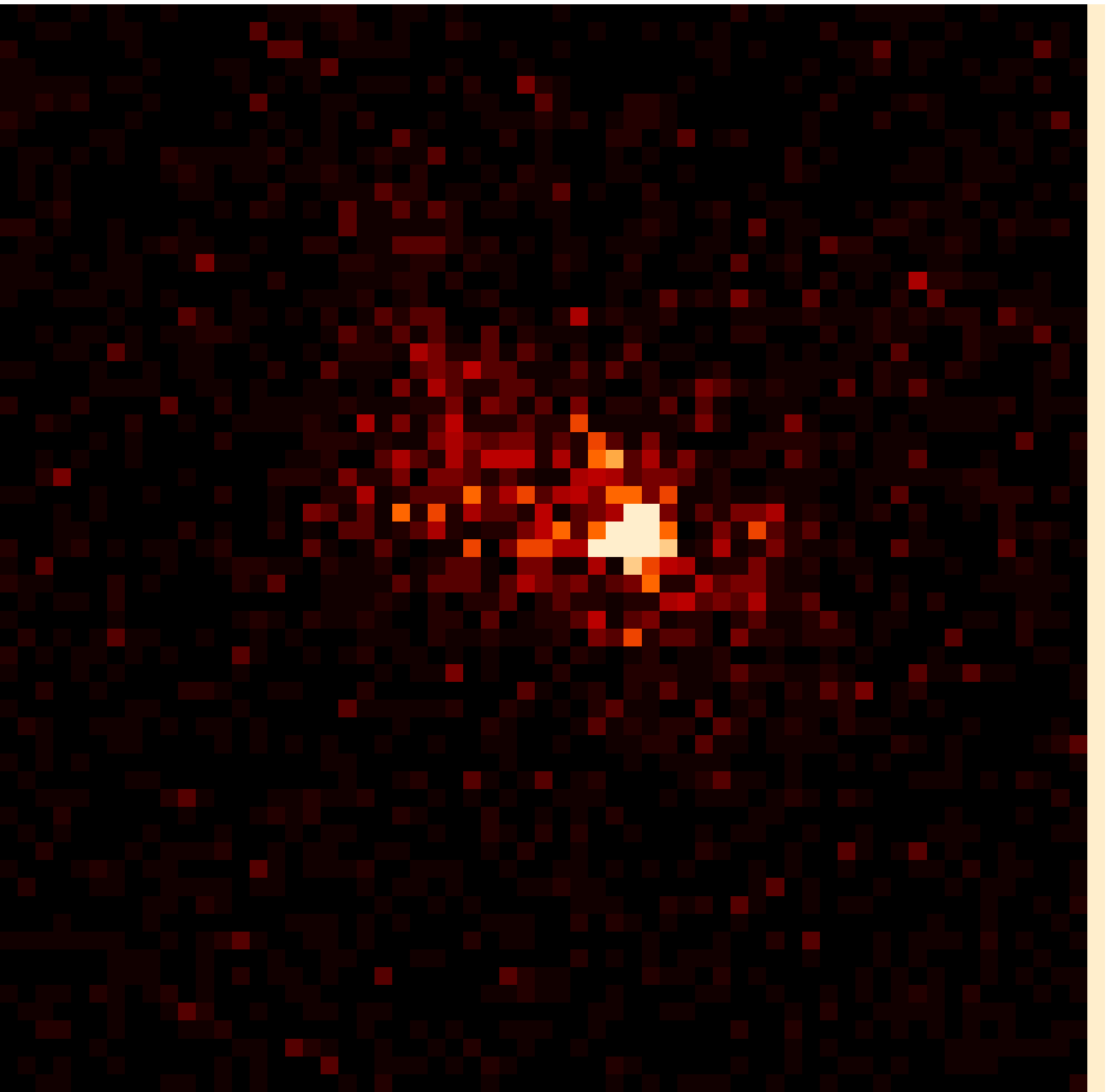}
   \caption{The $2-10$~keV images of the $5'\times5'$ region around the galactic nucleus obtained from MOS events integrated in the 1000~s before the flare (left) and in the 1000~s including the flare (right). Pixels were rebinned to a size of $5.5''\times5.5''$. Sgr~A* position is right in the middle of the central bright pixel visible in the flare image (right).}
   \label{fig:images}
\end{center}
\end{figure}

This work focus on the search for variability of a central point source at the Sgr~A* position. To optimize the signal to noise ratio and considering the $15''$ half power diameter of the XMM-Newton point spread function, we extracted and analyzed counts from the $10''$ radius region centered on Sgr~A*. The observed count rate is stable around 0.08~cts~s$^{-1}$ till the last 900~s (fig.~\ref{fig:lightcurve}). Then\, it\, gradually\, increases\, to\, reach\, 0.24~cts~s$^{-1}$, $\approx 7\sigma$ over the average value before the flare. This increase is not detected in the count rate extracted from a region far from the source. Images integrated during the 1000~s before the flare and during the last 1000~s including the flare, clearly show the brightening of a central source (fig.~\ref{fig:images}). An analysis of the flaring source location shows that it is compatible with Sgr~A* within $1.5''$ (i.e. within the residual systematic uncertainties). The nearest X-ray point source identified by Chandra and associated with the IR and radio object IRS~13, is at an angular distance of $4''$. We conclude therefore that this flare is associated with Sgr~A*.

\begin{figure}
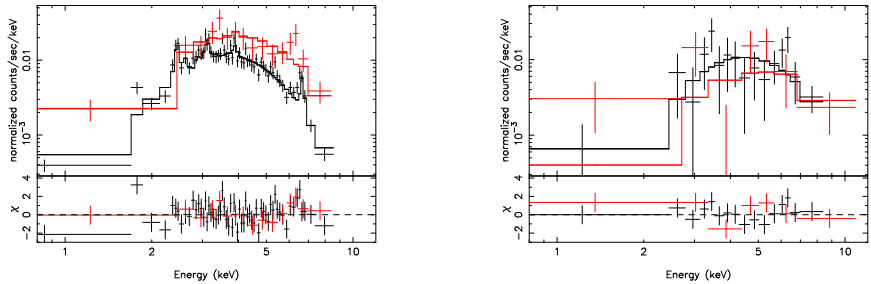

\begin{center}
   \includegraphics[width=4.9cm]{brion2_fig3a.eps}
   ~~~~~~~~~~~~
   \includegraphics[width=4.9cm]{brion2_fig3b.eps}
   \caption{Left: MOS count spectra from the region of $10''$ radius around Sgr~A* before the flare (black) and during the 900~s flare (red). Right: count spectra from MOS (black data point set) and PN (red data point set) data, extracted from the same region during the flare after subtraction of the non flaring spectra.}
   \label{fig:spectra}
\end{center}
\end{figure}

The spectra extracted from the same region as for the lightcurves before and during the flare were analyzed using a composite spectral model and fitted para\-me\-ters determined with the Chandra data (thermal and point sources components fixed, free power law for Sgr~A*) (fig.~\ref{fig:spectra}). The best fit gives a slope of $\alpha=0.9\pm0.5$ for Sgr~A* power law during the flare. Another spectral fit was made using the data before the flare as background component for the data during the flare, and we obtained $\alpha=0.7\pm0.6$ similar to the first fit. In spite of the large uncertainties the flare spectrum appears harder than the one measured with Chandra during the quiescent period. The flare luminosity we obtained (averaged over 900~s at 8~kpc) is $L_\mathrm{X} \mathrm{[2-10\ keV]}=(3.8\pm0.7)\times10^{34}$~erg~s$^{-1}$.

\section{Discussion}

The XMM-Newton discovery of a new X-ray flare of Sgr~A* confirms and strengthens the results obtained by Chandra. On September 2001, XMM-Newton observed the first part of a flare very similar to the first bright Chandra flare. The Sgr~A* flux increased by a factor 30 in 900~s with a hard spectral slope and a peak luminosity of $L_\mathrm{X} \mathrm{[2-10\ keV]}=6\times10^{34}$~erg~s$^{-1}$. This detection of another such a flare indicates that these events are not rare. From the total XMM-Newton and Chandra observation time (101~ks) the duty cycle of such an event (11~ks flare) is 0.11, i.e. one per day. Many observations of Sgr~A* have been made in radio and they have never shown a variability amplitude higher than a factor 2. The large X-ray flares don't seem associated to comparable radioflares, which is a constraint on the emission models for this source. Since this detection new observations have been made that confirm our conclusion. A Chandra monitoring of Sgr~A* in 2002 provided a rate of $\sim1.2\pm0.4$~flares/day without observed correlation at other wavelengths. XMM-Newton observed another flare in October 2002 (Porquet et al. 2003) very different from those reported before: it is brighter, shorter and softer and provides therefore further strong constraints for the theory.

Observations of the powerful stellar winds from the IR~16 stellar cluster give an expected accretion luminosity in the $2.6\times10^6\ \mathrm{M}_\odot$ central BH of $L_\mathrm{A}=1.8\times10^{43}$~erg~s$^{-1}$. But the maximal Sgr~A* quiescent bolometric luminosity is $5\times10^{36}$~erg~s$^{-1}$. The Bondi-Hoyle or Advection Dominated Accretion Flow (ADAF) models have tried to explain this poor accretion efficiency in the past (Melia \& Falcke 2001), but new or refined models are now necessary to explain the observed X-ray flares. The Bondi-Hoyle spherical accretion with sub-equipartition magnetic field can be associated to a compact hot Keplerian flow inside the circularization region (at $< 10\ R_\mathrm{S}$) (Liu \& Melia 2002). This model predicts a thermal synchrotron emission in the sub-mm band by the hot electrons in the disk and a Synchrotron Self Compton (SSC) emission dominating the X-ray band. An increase in accretion rate can result in bremsstrahlung, instead than SSC, to dominate the high energy emission and this is compatible with the flare hard spectrum we observed.  

Another set of models assume the presence of a non-thermal population of particles (either in a jet or in the flow) while the accretion flow is regulated by an ADAF disk where convection or outflows reduce the effective accretion rate (Markoff et al. 2001, Yuan et al. 2003). The sub-mm bump is produced by thin synchrotron emission of relativistic electrons at the base of a jet close to the BH, while synchrotron emission along the jet give rise to the radio spectrum. SSC is again responsible for the X-ray emission but in this model the flares must be produced by an increase of the electron temperature and not by a change in the accretion rate otherwise the flare X-ray spectrum would be soft and correlated to a radio flare, contrary to our results. Another interesting possibility for this model is that the synchrotron emission extends to the X-rays and dominates the high energy emission in flaring phases. The predicted $\gamma$-ray emission could be seen with INTEGRAL. Using the spectral shape from this XMM-Newton observation and the intensity at the peak of the Chandra flare, the expected signal to noise ratio for a 10~ks flare at $20-60$~keV is $5-10\ \sigma$ with the IBIS telescope.

Simultaneous X-ray and radio/sub-mm/IR/$\gamma$ observations of flares are therefore very important to understand the real nature of the main emission process in the massive BH of our Galaxy.


\end{document}